\documentclass[useAMS,fleqn,usenatbib]{mn2e}
\usepackage{amssymb,amsmath}
\usepackage{graphics}
\usepackage{tabularx}

\def\eqref#1{equation~(\ref{#1})}

\def\binom#1#2{{#1 \choose #2}}
\def\eoq{\mathrm{q}}
\def\eop{\mathrm{p}}

\def\fm{\mathrm{F}_{\rm m}}
\def\rv{V_{\rm r}}
\def\rvn#1{\rv^{(#1)}}

\title[Radial velocity analysis of multiple planetary systems]%
{Analysis of radial velocity variations in multiple planetary systems}
\author[A. P\'al]{
\noindent Andr\'as P\'al\thanks{E-mail: \texttt{apal@szofi.net}} \\
Konkoly Observatory of the Hungarian Academy of Sciences, 
	Konkoly Thege Mikl\'os \'ut 15-17,
	H-1121 Budapest, Hungary \\
Department of Astronomy, Lor\'and E\"otv\"os University,
	P\'azm\'any P. st. 1/A,
	Budapest H-1117, Hungary}

\begin{document}

\date{Accepted \dots. Received \dots; in original form \dots}

\pagerange{\pageref{firstpage}--\pageref{lastpage}} \pubyear{2009}

\maketitle

\label{firstpage}

\begin{abstract}

The study of multiple extrasolar planetary systems has the opportunity
to obtain constraints for the planetary masses and orbital inclinations
via the detection of mutual perturbations. The analysis of precise 
radial velocity measurements might reveal these planet-planet interactions
and yields a more accurate view of such planetary systems. Like in the
generic data modelling problems, a fit to radial velocity data 
series has a set of unknown parameters of which parametric derivatives
have to be known by both the regression methods and the estimations
for the uncertainties. In this paper an algorithm is described that
aids the computation of such derivatives in case of when planetary
perturbations are not neglected. The application of the algorithm
is demonstrated on the planetary systems of HD~73526, HD~128311 and HD~155358.
In addition to the functions related to radial velocity 
analysis, the actual implementation of the algorithm contains 
functions that computes spatial coordinates, velocities and barycentric 
coordinates for each planet. These functions aid the joint analysis
of multiple transiting planetary systems, transit timing and/or duration
variations or systems where the proper motion of the host star is also 
measured involving high precision astrometry. The practical implementation
related to the above mentioned problems features functions that make
these kind of investigations rather simple and effective. 

\end{abstract}



\section{Introduction}
\label{sec:introduction}

As of this writing, $36$ multiple planetary systems are known around 
main sequence stars. Most of these systems bear 
two detected planets, $8$ of them have $3$ and $2$ of them have $4$ 
planets while the star 55~Cnc has 5 
companions\footnote{See e.g. \texttt{http://exoplanet.eu} for an
up-to-date list.}. With the exception of HR~8799 \citep{marois2008}, 
all of the detections 
are based on or confirmed by the measurements of the 
radial velocity (RV) variations 
of the host stars\footnote{The planetary system around HR~8799 with
3 confirmed planets has been detected by direct imaging.}.
The planet HAT-P-13b \citep{bakos2009} also transits its host star.
The detection of this planet was based on transit photometry
while the second companion in this system has been revealed by
radial velocity measurements.
In general, analysis of RV variations constrains the mass ($m$) of planets 
by a lower limit. Namely, only the quantity $m\sin i$ is determined by
RV data where $i$ is the orbital inclination (relative to
the tangential plane of sky). With the exception of transiting planets
and systems were the spatial motion of the host star is detected
via astrometry, there
is no direct evidence for the actual value of the orbital inclination
(and therefore the mass of the planet). In case of transiting systems,
inclinations are constrained by measuring the impact parameter from
the light curves \citep[see e.g.][]{pal2010}, while astrometry
yields not only the inclination but the orientation of the orbital plane
as well \citep{bean2009,benedict2010,mcarthur2010}. 
The planetary system around GJ~876 
is the only known one where mutual inclination is also detected
with a 2-$\sigma$ confidence \citep{bean2009}.
A great advantage of multiple planetary systems is the possibility 
of detecting mutual perturbations via the deflection of RV values
from the purely Keplerian solution \citep[see e.g.][]{laughlin2001}. 
Therefore, precise analysis
of accurate RV series may yield to an acceptable constraint for
both the inclination and the planetary masses. Additionally, 
planet-planet interactions depend on the mutual inclinations, thus
more complex models (such as non-coplanar orbits) for the whole 
planetary systems can be investigated. 

Like in the majority of data modelling problems, RV variations
(of the host star) in single or multiple planetary systems 
are modelled with a function of a few external (unknown) parameters.
These parameters include the
orbital elements and the masses of the planets as well as the barycentric 
velocity of the host star. For most of the regression methods
involved in data modelling 
\citep[see e.g. the Levenberg-Marquard algorithm,][]{press1992},
and for the analytic estimation of the covariances, uncertainties and
correlations of the model parameters 
\citep[see e.g.][]{finn1992,pal2009}, the partial derivatives of
the model functions (with respect to the model parameters) 
have to be known in advance. 
The simplest way of RV curve modelling does not take into account the
mutual interactions between the planets and characterizes the
observed RV variations as a sum of independent Keplerian models.
\cite{wright2009b} describes an algorithm detailing the efficient
computation of the parametric derivatives of RV model functions where the
mutual planetary perturbations are neglected. The main objective of this
paper is to present an algorithm that calculates 
parametric derivatives when the planet-planet interactions
are also taken into account. 

The structure of the paper is as follows. In the next section, 
we describe the mathematical tools used to construct the algorithm itself
(including the discussion of optimal orbital parameterization,
and the numerical integration). In Section~\ref{sec:implementation} 
the practical implementation is detailed while in Section~\ref{sec:applications}
we demonstrate the usage of the algorithm for three specific multiple
planetary systems. The results are summarized in the last section.

\section{Orbital parameterization and numerical integration}
\label{sec:algorithm}

In this section we summarize the conventions involved in the orbital
parameterizations (as used throughout this paper), the algorithms
used for numerical integrations and the methods applied in the
calculations of parametric derivatives of the radial velocity
model functions.

\subsection{Spectroscopic and Keplerian orbital elements}
\label{sec:speckepelements}

The measured radial velocity variations of the host star determine
the \emph{spectroscopic} orbital elements of the planetary companion(s). 
In the case of a system with a single planet, there are $6$ of such parameters:
the period, $P$, the time of periastron passage, $T_{\rm p}$, 
eccentricity, $e$, argument of periastron, $\omega$, 
the semi-amplitude of the RV variations, $K$,
and the zero-point (or the mean) radial velocity of the central body, $\gamma$.
The above set of orbital parameters is the most widely used in the
literature, however, it is not the best choice for our purposes because of the 
following reasons. First, for nearly circular orbits the 
argument of pericenter is not so well constrained and for exactly
circular orbits it cannot be defined at all. Second, the
time of pericenter passage is also purely constrained for orbits
with small eccentricities and not defined for $e=0$. In order
to avoid such ambiguities, one can use the Lagrangian orbital elements
$k=e\cos\omega$ and $h=e\sin\omega$ for the parameterization of the
shape of the orbit and use one of the following quantities instead of
$T_{\rm p}$: the mean longitude $\lambda\equiv\lambda(E_0)$ or the 
orbital longitude $\varphi\equiv\varphi(E_0)$ for a certain fixed epoch $E_0$, 
or the moments $T_{\lambda_0}$ 
or $T_{\varphi_0}$ when the mean longitude or the orbital longitude have
a certain fixed value of $\lambda_0$ or $\varphi_0$, respectively. All 
of the above four quantities are well-defined for circular and nearly circular
orbits. The element $T_{\varphi_0}$ is widely used in the case of 
transiting planets since due to the definition of the 
alignment of the reference frame, transits occur at $\varphi_0=\pi/2$. 
In the case of multiple planetary systems where the mutual
interactions are not negligible, the usage $T_{\lambda_0}$ or $T_{\varphi_0}$ 
is not the best choice since in practice these imply different
epochs for the distinct planets. Throughout this paper, we use 
the mean longitude $\lambda$ as the primary orbital element to characterize
the phase of the orbital motion. The conversion between $T_{\rm p}$ 
and $\lambda$ is rather simple, namely
\begin{equation}
\lambda = \frac{2\pi}{P}\left(E_0-T_{\rm p}\right)+\omega.
\end{equation}
Since the mean longitude at an arbitrary moment $t$ is 
$\lambda(t)=n(t-E_0)+\lambda$, i.e. it is a linear function of 
the mean motion $n=(2\pi)/P$ and the $\lambda$, we prefer $n$ to $P$.
There is an additional benefit using $n$ for characterizing orbital periods:
simple error propagation estimations yield that for a given RV semi-amplitude,
the uncertainty of $n$ is independent for $n$ itself. In other words, 
in a multiple planetary systems where planets have masses with the same 
magnitude, one can expect that the obtained uncertainties are also roughly
the same. This is also confirmed by the demonstration analysis presented
here (see Sec.~\ref{sec:orbitalfits}).

In order to interpret the spectroscopic orbital elements, these
should be converted into Keplerian parameters. For any 
type of orbits, the orbital eccentricity, argument of periastron,
and the mean longitude are interpreted in the same way. 
In the case of planar orbits, semimajor axis $a$ of the orbit and 
the mass of the planet $m$ are derived 
from the mean motion $n$, and
the normalized RV semi-amplitude $\mathcal{K}=KJ$, where
$J\equiv\sqrt{1-e^2}=\sqrt{1-k^2-h^2}$.
Let us denote the mass of the central star by $\mathcal{M}$.
Using Kepler's Third Law and the barycentric velocity of the central 
body, namely
\begin{equation}
a^3n^2=G(\mathcal{M}+m)
\end{equation}
and 
\begin{equation}
\mathcal{K}=an\frac{m}{\mathcal{M}+m}.
\end{equation}
we obtain for $Gm$ and $a$:
\begin{eqnarray}
Gm &=& (G\mathcal{M})\fm\left(\frac{\mathcal{K}^3}{G\mathcal{M}n}\right), \\
a & =& n^{-2/3}\left\{G\mathcal{M}+Gm\right\}^{1/3} = \\
 & = & n^{-2/3}\left\{G\mathcal{M}\left[1+\fm\left(\frac{\mathcal{K}^3}{G\mathcal{M}n}\right)\right]\right\}^{1/3}. \nonumber
\end{eqnarray}
Here $\fm(\alpha)$ denotes the solution of the equation
\begin{equation}
\alpha=\frac{x^3}{(1+x)^2}.
\end{equation}
If $\alpha\ge 0$, the above equation always has a unique solution.
Moreover, if $\alpha>0$, the $\fm(\alpha)$ function behaves analytically and
its derivative is 
\begin{equation}
\frac{{\rm d}\fm(\alpha)}{{\rm d}\alpha}=\frac{[1+\fm(\alpha)]\fm(\alpha)}{\alpha[3+\fm(\alpha)]}.
\end{equation}
The well-known proportionality $\mathcal{K}\propto Gm$ is a direct
consequence of $\fm(\alpha)\approx \alpha^{1/3}$ for $\alpha\ll 1$.
We note that in the case of 
spatial orbits when the orbital inclination $i$
differs from $90^\circ$, the gravitational parameter and the semimajor
axis are calculated as above, but the normalized RV semi-amplitude
is $\mathcal{K}=KJ(\sin i)^{-1}$.

Once the Keplerian orbital elements are derived, the planar 
coordinates and velocity vector components 
of the planet can be calculated as
\begin{eqnarray}
\binom{x}{y} & = & a\left[\binom{c}{s}+\frac{p}{1+J}\binom{+h}{-k}-\binom{k}{h}\right], \label{eq:xyanal} \\ 
\binom{\dot x}{\dot y} & = & \frac{an}{1-q}\left[\binom{-s}{+c}+\frac{q}{1+J}\binom{+h}{-k}\right]. \label{eq:vxvyanal}
\end{eqnarray}
Here $q=e\cos E$, $p=e\sin E$, $c=\cos(\lambda+p)$ and $s=\sin(\lambda+p)$,
while $E$ denotes the eccentric anomaly. As it is shown by \cite{pal2009},
the quantities $q$, $p$, $c$ and $s$ are analytic functions of the
mean longitude and the Lagrangian orbital elements $(k,h)$. The 
observed radial velocity of the central body is then
\begin{equation}
\rvn{1}=-\frac{m}{\mathcal{M}+m}(o_x\dot x+o_y\dot y), \label{eq:rv1body}
\end{equation}
where the unity vector $(o_x,o_y)$ shows the direction of the
observer. Due to the historical definition of the reference
frames used in the astrophysics of binary stars and extrasolar
planets, this vector is fixed to be $(o_x,o_y)=(0,1)$.

\subsection{Lie-integration}

In this short section we summarize the properties of the numerical 
integration of ordinary differential equations named Lie-integration.
The main feature of the numerical integration based on the Lie-series
\citep{grobner1967,hanslmeier1984,eggl2010} is that the solution of the
differential equation
\begin{equation}
\dot x_i=f_i(\mathbf{x}), \label{diffeq}
\end{equation}
is approximated by its Taylor series up to a finite order. The 
coefficients for this power series expansion
are generated by the Lie-operator
\begin{equation}
L_0 := \sum\limits_{i} f_iD_i, \label{lieopdef}
\end{equation}
(where $D_i\equiv\frac{\partial}{\partial x_i}$)
with which the solution of \eqref{diffeq} can be written as
\begin{equation}
\mathbf{x}(t+\Delta t) = \exp\left(\Delta t\cdot L_0\right)\mathbf{x}(t) \label{liesol1}
\end{equation}
where
\begin{equation}
\exp\left(\Delta t\cdot L_0\right)=
\sum\limits_{k=0}^\infty \frac{\Delta t^k}{k!}L_0^k=
\sum\limits_{k=0}^\infty \frac{\Delta t^k}{k!}
\left(\sum\limits_{i} f_iD_i\right)^k. \label{liesol2}
\end{equation}
The method of Lie-integration is the finite approximation 
of the sum in the right-hand side of 
\eqref{liesol2}, up to the order of $M$. Thus, the solution after $\Delta t$
time is approximated by 
\begin{equation}
\mathbf{x}(t+\Delta t)\approx\left(\sum\limits_{k=0}^{M} \frac{\Delta t^k}{k!}L_0^k\right)
\mathbf{x}(t)=\sum\limits_{k=0}^{M} \frac{\Delta t^k}{k!}\left(L_0^k\mathbf{x}(t)\right).\label{lienumint}
\end{equation}
Supposing the coefficients $L_0^k\mathbf{x}(t)$ are computed, the numerical
integration itself is straightforward. In practice, the values
of these coefficients are computed involving recurrence relations,
i.e. for $n\ge 0$, the $L_0^{n+1}\mathbf{x}(t)$ terms are evaluated
using the previously calculated $L_0^k\mathbf{x}(t)$ ($0\le k \le n$)
coefficients. For each particular problem, the recurrence relations
must be derived properly. For the general $N$-body problem, these
relations are presented in \cite{hanslmeier1984} while the relations
in the cases when the reference frame is fixed to one of the 
bodies are shown in \cite{pal2007}.

The computational cost (CPU time) required by the calculation
of the $L_0^k\mathbf{x}(t)$ coefficients is definitely larger 
than the time of evaluating \eqref{lienumint} \citep[see][]{pal2007}.
Thus, a great advantage of the Lie-integration is that the stepsize $\Delta t$
can be altered after the coefficients are evaluated without much of
additional cost, and therefore an effective integration method can be
implemented with an adaptive stepsize. Moreover, there is an
availability of an alternate way of adaptive integration, namely
if the stepsize $\Delta t$ is kept fixed, the integration order $M$ 
can also dynamically be increased until we reach the desired 
precision \citep[see also][]{eggl2010}.
More details about the practical implementation are found 
in Sec.\ref{subsec:adaptiveintegration}.

\subsection{Motion in the reference frame of one of the bodies}
\label{sec:fixedequations}

In several applications, such as in the description of 
a planetary system or in perturbation theory,
the equations of motion are transformed 
into a reference frame whose origin coincides with one of the bodies.
Practically, this is the body with the largest mass, i.e.
in a planetary system it is the host star. 
Let us define the central body as the body with the index of $i=0$.
Altogether we have $1+N$ bodies, where the other ones are indexed
by $i=1,\dots,N$. 
Let us use the relative (non-inertial)
coordinates $r_{im}$ and velocities $w_{im}$ (where the second index $m$
refers to the spatial dimension, i.e. $m=1$ is for the $x$ coordinate,
$m=2$ is the $y$ coordinate and in non-planar problems, $m=3$ refers
to the $z$ coordinate). The equations of motion
in a compact form are
\begin{eqnarray}
\dot r_{im} & = & w_{im}, \label{fixednb}\\
\dot w_{im} & = & -G(\mathcal{M}+m_i)\phi_i r_{im} - \nonumber \\
	& & - G\sum\limits_{j=1, j\ne i}^{N}m_j\left[\phi_{ij}A_{ijm}+\phi_j r_{jm}\right], \label{fixednb2}
\end{eqnarray}
where $A_{ijm}$ is the $m$th component of the vector pointing from
the body $j$ to body $i$,
\begin{equation}
A_{ijm} :=  r_{im}-r_{jm},
\end{equation}
$\rho_i$ and $\rho_{ij}$ denotes the distances from the central
body and the mutual distances, respectively:
\begin{eqnarray}
\rho_i & := & \rho_{0i} = \rho_{i0} = \sqrt{\sum\limits_m r_{im}r_{im}}, \\
\rho_{ij} & := & \sqrt{\sum\limits_m A_{ijm}A_{ijm}},
\end{eqnarray}
and $\phi_i$ and $\phi_{ij}$ are defined as the reciprocal cubic distances,
\begin{eqnarray}
\phi_{ij} & := & \rho_{ij}^{-3},  \\
\phi_i & := & \rho_i^{-3}. 
\end{eqnarray}
Note that the quantities $\rho_i$ and $\rho_{ij}$, like so 
$\phi_i$ and $\phi_{ij}$ are distinguished only by the number of their indices.
Without going into details, we present the recurrence relations of
the Lie-derivatives, including the linearized variables 
in Appendix~\ref{appendix:fixednb}. 

Using the notations introduced above, the observed radial velocity
is computed as
\begin{equation}
V_{\rm r}^{(N)}=
\sum\limits_m\left(\frac{\sum\limits_{i=1}^{N}m_i w_{im}}{\mathcal{M}+\sum\limits_{i=1}^{N}m_i}\right)o_m,\label{eq:rvnbody}
\end{equation}
where the unity vector $\mathbf{o}\equiv o_m=(o_x,o_y)$ 
or $o_m=(o_x,o_y,o_z)$ defines the direction of the
observer (see also equation~\ref{eq:rv1body}).

\subsection{Parametric derivatives}
\label{subsec:paramderivatives}

In the analysis of radial velocity variations in a multiple planetary
system, the parametric derivatives of the RV curve have to be computed
at a certain moment $t$ with respect to the initial orbital
elements (at $t=E_0$). Let us define an arbitrary quantity $Q$
which depends only on the solution of the ordinary differential
equation (\ref{diffeq}), $Q\equiv Q(\mathbf{x}(t))\equiv Q(\mathbf{x})$.
The parametric derivatives of the quantity $Q$ with respect to the 
initial conditions $\mathbf{x}^0\equiv \left.\mathbf{x}\right|_{t=0}$ 
can be computed in the following way. First, let us write
the linearized equations of \eqref{diffeq} as 
\begin{equation}
\dot{\xi}_k=\xi_{m}\frac{\partial f_k(\mathbf{x})}{\partial x_m}.\label{eq:simplelinearized}
\end{equation}
(Here and in the following we use the implicit summation notation wherever
it is unambiguous.) The variables $\xi_k$ denote the so-called linearized
variables for one particular initial condition. Second, due to the
linear property, with the solution of the \emph{full linearized} equations
\begin{equation}
\dot{\mathcal{Z}}_{\ell k}=\mathcal{Z}_{\ell m}\frac{\partial f_k(\mathbf{x})}{\partial x_m}\label{eq:fulllinearized},
\end{equation}
one can compute the solution of \eqref{eq:simplelinearized} for any arbitrary
initial conditions $\xi_k^0$, namely:
\begin{equation}
\xi_k(t)=\mathcal{Z}_{k\ell}(t)\xi_{\ell}^0
\end{equation}
if the respective initial conditions of \eqref{eq:fulllinearized} are
\begin{equation}
\left.\mathcal{Z}_{\ell k}\right|_{t=0}=\delta_{\ell k}=
\left\{\begin{tabular}{l}1 ~ if ~ $\ell=k,$\\0 ~ if ~ $\ell\ne k.$\end{tabular}\right.
\end{equation}
Finally, without going into the details, it can be shown that the 
partial derivatives of $Q(t)$ with respect to the initial conditions 
$\mathbf{x}^0\equiv \left.\mathbf{x}\right|_{t=0}$ is
\begin{equation}
\frac{\partial Q(t)}{\partial x^0_\ell}=\mathcal{Z}_{\ell k}(t)\frac{\partial Q}{\partial x_k}. \label{eq:qpartial1}
\end{equation}
where $\mathcal{Z}_{\ell k}(t)$ represents the solution of \eqref{eq:fulllinearized}
at the instance $t$. 
If the initial conditions $\mathbf{x}^0$ are defined with an alternative
parameterization, i.e. $\mathbf{x}^0=\mathbf{x}^0(\hat{\mathbf{x}})$,
the parametric derivatives of the quantity $Q$ with respect
to the $\hat{\mathbf{x}}$ are calculated involving the chain rule, 
namely
\begin{equation}
\frac{\partial Q}{\partial \hat{x}_\ell}=\frac{\partial x^0_\ell}{\partial\hat{x}_m}\mathcal{Z}_{mk}\frac{\partial Q}{\partial x_k}. \label{eq:qpartial2}
\end{equation}
In the analysis of RV data series, $\hat{\mathbf{x}}$ represents
the set of spectroscopic orbital elements -- including the (normalized)
RV semi-amplitude, 
$\hat{\mathbf{x}}\equiv(\mathcal{K}_i,n_i,\lambda_i,k_i,h_i)$ --,
$\mathbf{x}^0$ represents the spatial coordinates,
velocities\footnote{See also 
Sec.~\ref{sec:fixedequations} for further details about the notations
used in the description of multiple planetary systems.}
and the gravitational parameters of the planets,
while $Q\equiv V_{\rm r}^{(N)}$, the observed radial velocity. 
In practice, the partial derivatives $\frac{\partial x^0_\ell}{\partial\hat{x}_m}$
can be computed using the formulae presented in Appendix~\ref{appendix:partial},
while the computation of the terms $\frac{\partial Q}{\partial x_k}$
is relatively simple since in the equation for the radial 
velocity (see equation~\ref{eq:rv1body} or later in
Sec.~\ref{sec:fixedequations}, equation \ref{eq:rvnbody})
is a rational expression of two functions in that are linear 
with respect to both the coordinates and masses.

\subsection{Linearized equations}

As we have seen before (Sec.~\ref{subsec:paramderivatives}), 
linearized equations have to be solved in order to calculate the partial 
derivatives of an arbitrary
quantity (that depends on the solution of the original differential
equation) with respect to the initial conditions. As it has
been shown in \cite{pal2007}, using the same notations as above 
the Lie-derivatives of 
the partial linearized variables $\xi_k$ (see also 
the previous subsection) can be written as
\begin{equation}
L^n\xi_k=\xi_mD_mL^nx_k=\xi_mD_mL_0^nx_k.\label{eq:linliederiv}
\end{equation}
Obviously, this formula can be 
applied to obtain the solution for the full linearized form
(see equation~\ref{eq:fulllinearized}):
\begin{equation}
L^nZ_{\ell k}=Z_{\ell m}D_mL_0^nx_k. \label{fulllinlie}
\end{equation}
Thus, the solution for \eqref{fulllinlie} has to be 
substituted into \eqref{eq:qpartial1} or (\ref{eq:qpartial2}) in order
to obtain the partial derivatives of arbitrary quantities with respect to
the initial conditions. The complete set of recurrence relations for
the linearized problem is found in Appendix~\ref{appendix:fixednb}.


\begin{table*}
\caption{List of additional modes as implemented in 
the generic functions \texttt{nbrv\_2g\_N()} and \texttt{nbrv\_3g\_N()}. 
These functions have $4$ or $5$ additional parameters (in the respective
cases of $2$ and $3$ dimensional variants) comparing to the 
pure RV functions (\texttt{nbrv\_2d\_N()} and \texttt{nbrv\_3d\_N()}).
The first additional parameter is the ``mode flag'', $F$, an integer between 
$0$ and $3$. The second parameter is the body index, $k$, a non-negative 
integer less than or equal to $N$. The other $2$ or $3$ parameters
are the components of the $o_m$ vector. All of the values
computed by these functions are \emph{projected} coordinates or
velocities: the spatial vectors are multiplied by the $o_m$ components,
yielding a scalar product. Thus, in this table such derived
coordinates and velocities are referred as ``projected coordinates''
or ``projected velocities''.}\label{tab:genfunctmodes}
\begin{center}
\renewcommand{\arraystretch}{1.2}
\begin{tabularx}{170mm}{lllX}
\hline
\hline
Mode			& Body 			& Result	& Interpretation and typical usage	\\
($F$)			& index ($k$)		\\
\hline
$0$			& $0$			& $V^{(N)}_{{\rm r},m}o_m$	&
	Projected velocity of the barycenter with respect to the central body.
	In case of planetary systems, interpreted as the radial 
	velocity of the host star where the line-of-sight is defined
	by the $o_m$ vector. If $\mathbf{o}=(0,1)$ or $\mathbf{o}=(0,0,1)$,
	the results are equivalent with the results of \texttt{nbrv\_2d\_N()}
	and \texttt{nbrv\_3d\_N()}. 
	\\
$0$			& $1\le k\le N$		& $V^{(N,k)}_{{\rm r},m}o_m$	& 
	Independent components of the projected barycentric velocity.
	Although only the joint effect of all of the planets in the 
	planetary system can be measured by radial velocity variations, 
	the contribution of each planet to the final RV curve can be
	analyzed by this way. Due to the mutual perturbations, 
	these velocity components are not strictly periodic and 
	cannot be described only by the orbital elements of the respective 
	planet. 
	\\
$1$			& $0$			& $B^{(N)}_{{\rm r},m}o_m$	&
	Projected coordinates of the barycenter with respect to the 
	central body. In case of planetary systems, these coordinates
	describes the wobbling of the host star, as it might be detected 
	by precise astrometric measurements. 
	\\
$1$			& $1\le k\le N$		& $B^{(N,k)}_{{\rm r},m}o_m$	& 
	Independent components of the projected barycentric coordinates. 
	If the complementary inclination for a particular planet is 
	close to zero, the planet transits the host star, yielding 
	a small flux decrease that can be measured. In case of such transiting 
	planets, these coordinates determines the magnitude of the 
	transit timing variations due to light-time effects.
	\\
$2$			& $1\le k\le N$		& $w_{km}o_m$			&
	Projected spatial velocity of the body $k$. 
	For planets with a nearly edge-on orbit, the
	tangential acceleration of transiting planets is negligible
	at the time of the transits. These velocities well constrain
	the duration of these transits.
	\\
$3$			& $1\le k\le N$		& $r_{km}o_m$			&
	Projected spatial coordinates of the body $k$. These 
	coordinates constrain the shape of these
	possible transit light curves as well as the precise timings
	of the transits if an orbit is nearly edge-on.
	\\
\hline
\hline
\end{tabularx}
\end{center}
\end{table*}

\begin{table*}
\caption{Spectroscopic orbital elements
for the planetary systems 
HD~73526, HD~128311, and HD~155358.
These orbital elements have been derived from the
data available in the literature (see text for further references)
and have been used as an initial condition in the fits discussed
in this paper. The last column shows the number of available radial velocity
data points (the same as involved in the fits).}
\begin{center}
\renewcommand{\arraystretch}{1.2}
\begin{tabular}{lllrrrrrr}
\hline
\hline
System	& $E_0$ (BJD)	& $M_\star/M_\odot$ 
	& $\mathcal{K}_i\sin i$ (m/s)	& $n_i=2\pi/P_i$ (1/d) 
	& $\lambda_i$ (rad) &	$k_i=e_i\cos\omega_i$	& $h_i=e_i\sin\omega_i$ 
	& $N_{\rm RV}$	\\
\hline
HD73526	& $2,452,500$	& $1.08$
	& $70.0$	& $0.03360$ & $3.902$	& $-0.402$ & $+0.040$ &	$31$ \\
& &	& $61.4$	& $0.01620$ & $4.150$	& $-0.480$ & $-0.080$ \\
\hline
HD128311& $2,452,500$	& $0.84$
	& $64.6$	& $0.01370$ & $1.896$	& $-0.090$ & $+0.233$ &	$75$ \\
& &	& $75.1$	& $0.00677$ & $1.500$	& $-0.160$ & $-0.058$ \\
\hline
HD155358& $2,453,500$	& $0.87$ 
	& $34.6$	& $0.03222$ & $0.894$	& $-0.106$ & $+0.035$ &	$71$ \\
& &	& $14.1$	& $0.01185$ & $0.249$	& $+0.027$ & $-0.174$ \\
\hline
\hline
\end{tabular}
\end{center}
\label{tab:orbitalelements}
\end{table*}

\begin{table*}
\caption{Best-fit spectroscopic orbital elements
and their 1-$\sigma$ uncertainties for the planetary systems 
HD~73526, HD~128311, and HD~155358
derived by the MCMC algorithm under the assumption of coplanar 
orbits. The median for each probability distribution is treated as
a best-fit value. See text for further details.}
\begin{center}
\renewcommand{\arraystretch}{1.2}
\begin{tabular}{lrrrrrrr}
\hline
\hline
System	
	& $\mathcal{K}_i\sin i$ (m/s)	& $n_i=2\pi/P_i$ (1/d) 
	& $\lambda_i$ (rad) &	$k_i=e_i\cos\omega_i$	& $h_i=e_i\sin\omega_i$
	& $\gamma$ (m/s) & $\sin i$ \\
\hline
HD73526	
	& $65.9\pm3.8$	& $0.03359\pm0.00017$ & $3.855\pm0.046$	& $-0.404\pm0.045$ & $+0.069^{+0.047}_{-0.039}$ 
	& $-33.6\pm2.4$ & $0.82^{+0.18}_{-0.14}$ \\
	& $61.8\pm0.5$	& $0.01629^{+0.00018}_{-0.00015}$ & $4.108\pm0.076$	& $-0.503^{+0.044}_{-0.041}$ & $-0.046\pm0.046$ \\
\hline
HD128311
	& $49.1^{+9.2}_{-5.9}$	& $0.01364\pm0.00012$ & $1.599^{+0.219}_{-0.188}$	& $+0.035^{+0.112}_{-0.098}$ & $+0.337^{+0.071}_{-0.079}$ 
	& $0.9\pm2.1$	& $0.8^{+0.2}_{-0.5}$ \\
	& $73.8\pm3.1$		& $0.00663\pm0.00010$ & $1.396\pm0.046$			& $+0.142^{+0.093}_{-0.169}$ & $+0.063^{+0.065}_{-0.083}$ \\
\hline
HD155358
	& $31.18\pm0.34$		& $0.03225^{+0.00024}_{-0.00016}$ & $0.867\pm0.064$	& $-0.136\pm0.039$ & $+0.041^{+0.043}_{-0.047}$ 
	& $10.1\pm1.0$	& -- \\
	& $13.66^{+1.90}_{-1.58}$	& $0.01203^{+0.00029}_{-0.00033}$ & $0.161^{+0.147}_{-0.169}$	& $-0.065^{+0.091}_{-0.118}$ & $-0.138_{-0.179}^{+0.116}$ \\
\hline
\hline
\end{tabular}
\end{center}
\label{tab:orbitalfits}
\end{table*}

\section{Implementation}
\label{sec:implementation}

The algorithm presented in Section~\ref{sec:algorithm} has been implemented
as an add-on module for the regression analysis and data modelling 
program \texttt{lfit}\footnote{This program is available as a
part of the \texttt{libpsn} package, see 
\texttt{http://szofi.elte.hu/\~{ }apal/utils/libpsn/}.}
\citep[described briefly in][]{pal2009b} in a 
form of an ANSI C 
code. Since the application program interface (API) of 
\texttt{lfit} for this kind of dynamically loaded libraries is 
rather simple, the source code module can easily be modified for 
arbitrary purposes, such as inclusion for other kind of C programs
or another languages or programming environments that support 
linking of C modules (e.g., FORTRAN or IDL). The source code
is available from the web address 
\texttt{http://szofi.elte.hu/\~{ }apal/utils/astro/nbrv}.
A standalone implementation of the Lie-integrator code is also available
from the address \texttt{http://szofi.elte.hu/\~{ }apal/utils/astro/lieint},
with the same algorithmical features and with an easy user interface
for numerical integration and simple stability investigations.

In practice, this add-on module (named \texttt{nbrv.so} on most
of the UNIX systems or \texttt{nbrv.dylib} on OS/X) registers
the functions named \texttt{nbrv\_2d\_N()} 
and \texttt{nbrv\_3d\_N()} where $\texttt{N}\equiv N$
is the number of planets in the planetary 
systems\footnote{In the current implementation $N\le 8$, however,
the source code can easily be modified to increase the maximum 
number of planets.}. These
functions have $1+5N+1$ or $1+7N+1$ parameters, for the respective
cases for the planar (\texttt{nbrv\_2d\_N}) and 
spatial (\texttt{nbrv\_3d\_N}) problems. The first parameter is the 
central mass $\mathcal{M}$ (in Solar units), the 
following $N\times 5$ or $N\times 7$ parameters are the spectroscopic orbital parameters
$\mathcal{K}_i$ (in the units of ${\rm m/s}$), $n_i$ (in 
the units of ${\rm d}^{-1}$), $\lambda_i$ (in radians), $k_i$ and 
$h_i$. Furthermore, the spatial functions have two additional parameters:
the complementary orbital inclination, $\hat{\imath}\equiv 90^{\circ}-i$ 
and the argument of ascending node, $\Omega$, both angles are 
measured in radians\footnote{The complementary angle of the inclination 
is used for simplicity: the planar functions yields the 
same values as the spatial ones if these two additional 
parameters are set to zero.}.
All of these orbital elements are defined for a certain epoch of $E_0$ (BJD).
The last parameter is the time $\Delta t$ elapsed from the epoch $E_0$, i.e.
$\Delta t=t-E_0$. Here $i$ 
is the index for the actual planet, $1\le i \le N$.

\subsection{Adaptive integration}
\label{subsec:adaptiveintegration}

As mentioned earlier, one of the advantages of the Lie-integration
is the possibility of the implementation of a two-way adaptation,
by varying both the stepsize and the order of the integration. Since
the summation of the Taylor-coefficients in \eqref{lienumint} does not need so 
much computing time (compared to the evaluation of these coefficients),
the stepsize of the integration can easily be altered in order 
to reach the desired precision. In practice, the adaptive integration
is implemented as follows.
First, let us define a minimal and
maximal order of $M_{\rm min}$ and $M_{\rm max}$. 
It is easy to see
that an initial, nearly optimal stepsize for the integration is 
$\Delta t_0 \propto n_{\rm max}^{-1}$, where $n_{\rm max}$ is the
maximum of the mean motions appearing in the planetary system. In the case
of circular orbits, 
$\Delta t_0 = 0.8\,n_{\rm max}^{-1}$ is a good choice for $M\approx 20$
and for a relative precision of $\delta=2\cdot 10^{-16}$. 
If we allow a minimal and maximal order for the Lie-integration
($M_{\rm min}$ and $M_{\rm max}$, respectively),
the adaptive control of the stepsize and integration
order is done as:
\begin{description}
\item[\textbf{1.}] The terms $L^{k}_0\mathbf{x}$ in \eqref{lienumint} 
are evaluated using the appropriate recurrence relations and 
the summation is performed with a fixed value of $\Delta t$.
\item[\textbf{2.}] If the desired precision $\delta$ is reached before the
order of $M=M_{\rm min}$, $\Delta t$ is multiplied by the factor 
$M_{\rm max}/M_{\rm min}$
and re-compute the sum in \eqref{lienumint} (with the additional evaluation
of the necessary terms where $k>M$). This step is repeated until
$M <  M_{\rm min}$.
\item[\textbf{3.}] If the desired precision $\delta$ cannot be reached 
until the order of $M=M_{\rm max}$, divide $\Delta t$ by the factor
of $M_{\rm max}/M_{\rm min}$ and re-compute the sum in \eqref{lienumint}.
This step is repeated until $M_{\rm max} < M$.
\item[\textbf{4.}] If the desired precision is reached between the
orders of $M_{\rm min}$ and $M_{\rm max}$, accept the value of $\Delta t$
and proceed with the next step of the integration.
\end{description}
We have to note that this kind of two-way adaptive integration assures
that we definitely obtain the desired precision level \emph{without} 
the loss of computing time. In the case of more common integrators
(like Runge-Kutta or Bulirsch-Stoer methods),
the estimation of the accuracy is based on heuristics and it is not
checked by these algorithms that the expected precision is really obtained. 
If it turns out that the stepsize is too large (or other parameter of the 
integration should be changed), then a total re-computation is needed for
these algorithms and we definitely lose the computing time spent on the 
previous evaluations. 

In the actual implementation of the \texttt{nbrv} module, $\delta$ 
has been chosen by default to be the precision level of the IEEE 
double precision (64 bit) floating point number representation, 
that is $\delta=2\cdot 10^{-16}$. Although it is an extreme precision
compared to the implied and required precision level for the problem,
this precision implies that the whole set of functions implemented in 
the module can be treated as analytic functions without any side-effects. 
Additionally, this precision level ensures that there would not be any 
systematic distortions by varying the samples on the domain of investigations. 

\subsection{Properties}

In the following, we present some 
properties for these functions. For simplicity, let us denote 
by $\mathbf{S}_i$ the set of the spectroscopic orbital elements
$(n_i,\lambda_i,k_i,h_i)$ and define 
$V_{\rm r}^{(N)}(\cdot)=\texttt{nbrv\_N(.)}$. Since 
interaction between the planets is relatively small (comparing to
the gravitational force of the central star),
\begin{equation}
V_{\rm r}^{(N)}(G\mathcal{M},\mathcal{K}_1,\mathbf{S}_1,\dots,\mathcal{K}_N,\mathbf{S}_N,\Delta t) 
\approx \nonumber
\end{equation}
\begin{equation}
\hfill \approx \sum\limits_{i=1}^N V_{\rm r}^{(1)}(G\mathcal{M},\mathcal{K}_i,\mathbf{S}_i,\Delta t).\label{eq:rvsumapprox}
\end{equation}
It is easy to show that for fixed values of $\mathcal{K}_i$, 
the effect of mutual interactions decreases as the central mass 
increases, namely
\begin{equation}
\lim\limits_{G\mathcal{M}\to \infty}V_{\rm r}^{(N)}(G\mathcal{M},\mathcal{K}_1,\mathbf{S}_1,\dots,\mathcal{K}_N,\mathbf{S}_N,\Delta t) 
= \nonumber
\end{equation}
\begin{equation}
\hfill = \sum\limits_{i=1}^N V_{\rm r}^{(1)}(G\mathcal{M},\mathcal{K}_i,\mathbf{S}_i,\Delta t).\label{eq:masstoinfty}
\end{equation}
Similarly, scaling of the normalized RV semi-amplitudes yields the
same kind of equation:
\begin{equation}
\lim\limits_{\sin i\to \infty}(\sin i)V_{\rm r}^{(N)}\left(G\mathcal{M},\frac{\mathcal{K}_1}{\sin i},\mathbf{S}_1,\dots,\frac{\mathcal{K}_N}{\sin i},\mathbf{S}_N,\Delta t\right) 
= \nonumber
\end{equation}
\begin{equation}
\hfill = \sum\limits_{i=1}^N V_{\rm r}^{(1)}(G\mathcal{M},\mathcal{K}_i,\mathbf{S}_i,\Delta t).\label{eq:sinitoinfty}
\end{equation}
Although $|\sin i|\le 1$ for real orbits, either the above equation or
\eqref{eq:masstoinfty} can be used to formally decrease the level of 
interaction between the planets \citep{laughlin2001}. 
Additionally, the fit of
independent Keplerian orbits involving the right-hand side of 
\eqref{eq:rvsumapprox} yields good initial conditions for the
real problem. 

Additionally, using equations (\ref{eq:vxvyanal}) and 
(\ref{eq:rv1body}), $V_{\rm r}^{(1)}$ can also be written as
\begin{equation}
V_{\rm r}^{(1)}(G\mathcal{M},\mathcal{K}_1,\mathbf{S}_1,\Delta t)
=\frac{\mathcal{K}_1}{1-q}\left[\cos(\lambda+p)-\frac{k_1q}{1+J}\right]
\end{equation}
where $\mathbf{S}_1=(n_1,\lambda_1,k_1,h_1)$, 
$\lambda=n_1\Delta t+\lambda_1$, $p=\eop(\lambda,k_1,h_1)$,
$q=\eoq(\lambda,k_1,h_1)$ and $J=\sqrt{1-k_1^2-h_1^2}$. 
Obviously, $V_{\rm r}^{(1)}$ does not 
depend on $G\mathcal{M}$, therefore this argument is only a formal one.

As it is known from the literature of binary stars and 
hierarchical stellar systems, if $N=1$, variations in the argument 
of the ascending node have no observable effect and like so, for two planets,
radial velocity variations depend only on the difference 
$D=\Omega_2-\Omega_1$ (that also determines the mutual inclination of the
two orbit). In general, for all $N\ge 1$, we can state
\begin{equation}
\sum\limits_{k=1}^{N}\frac{\partial V_{\rm r}^{(N)}}{\partial\Omega_k}=0,
\end{equation}
that is also equivalent for the previously mentioned special cases
for $N=1$ and $N=2$. One should keep in mind these properties
while utilizing the \texttt{nbrv\_3d\_N()} functions for purely
radial velocity data.

\begin{figure*}
\begin{center}
\resizebox{55mm}{!}{\includegraphics{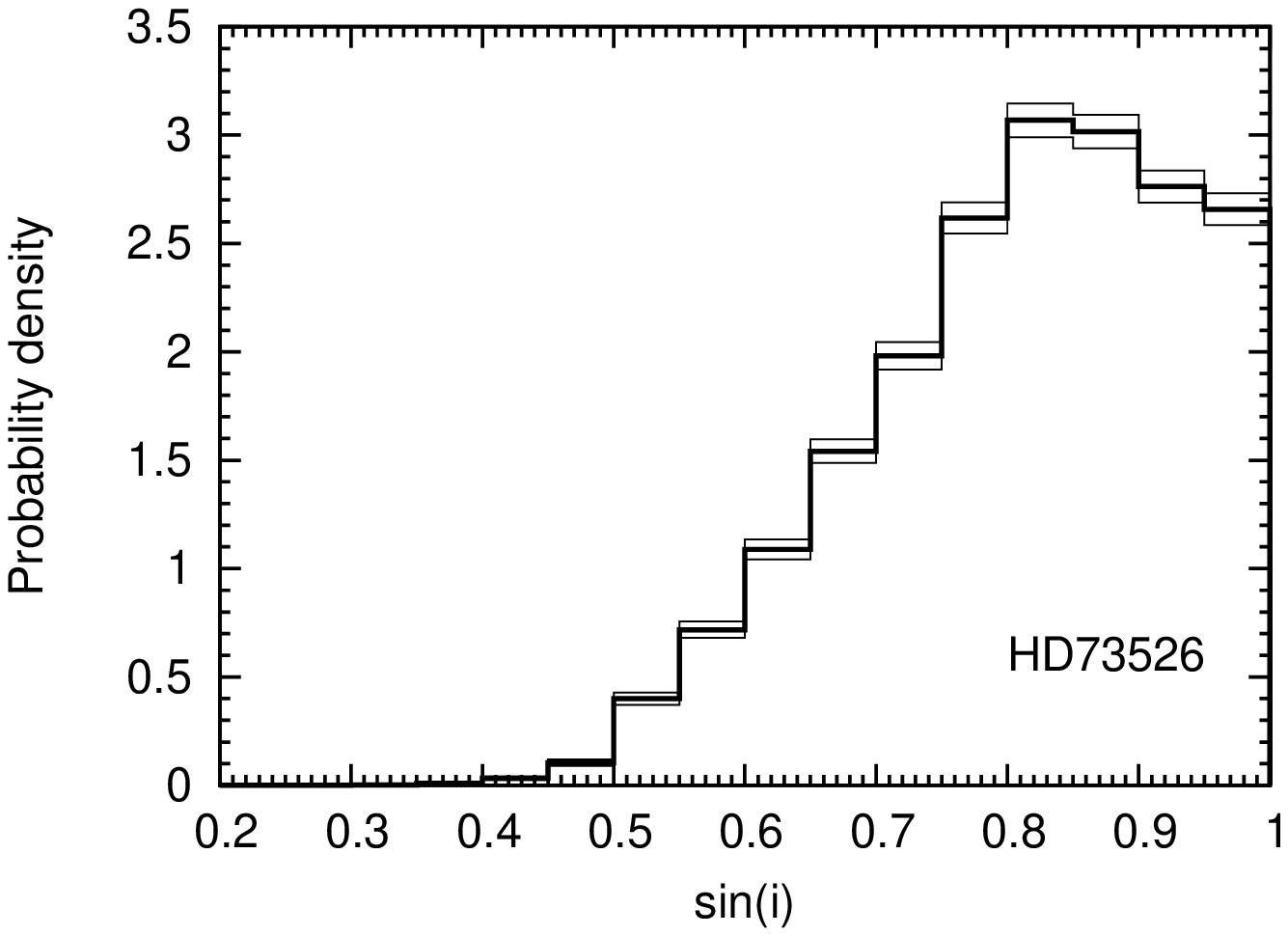}}%
\resizebox{55mm}{!}{\includegraphics{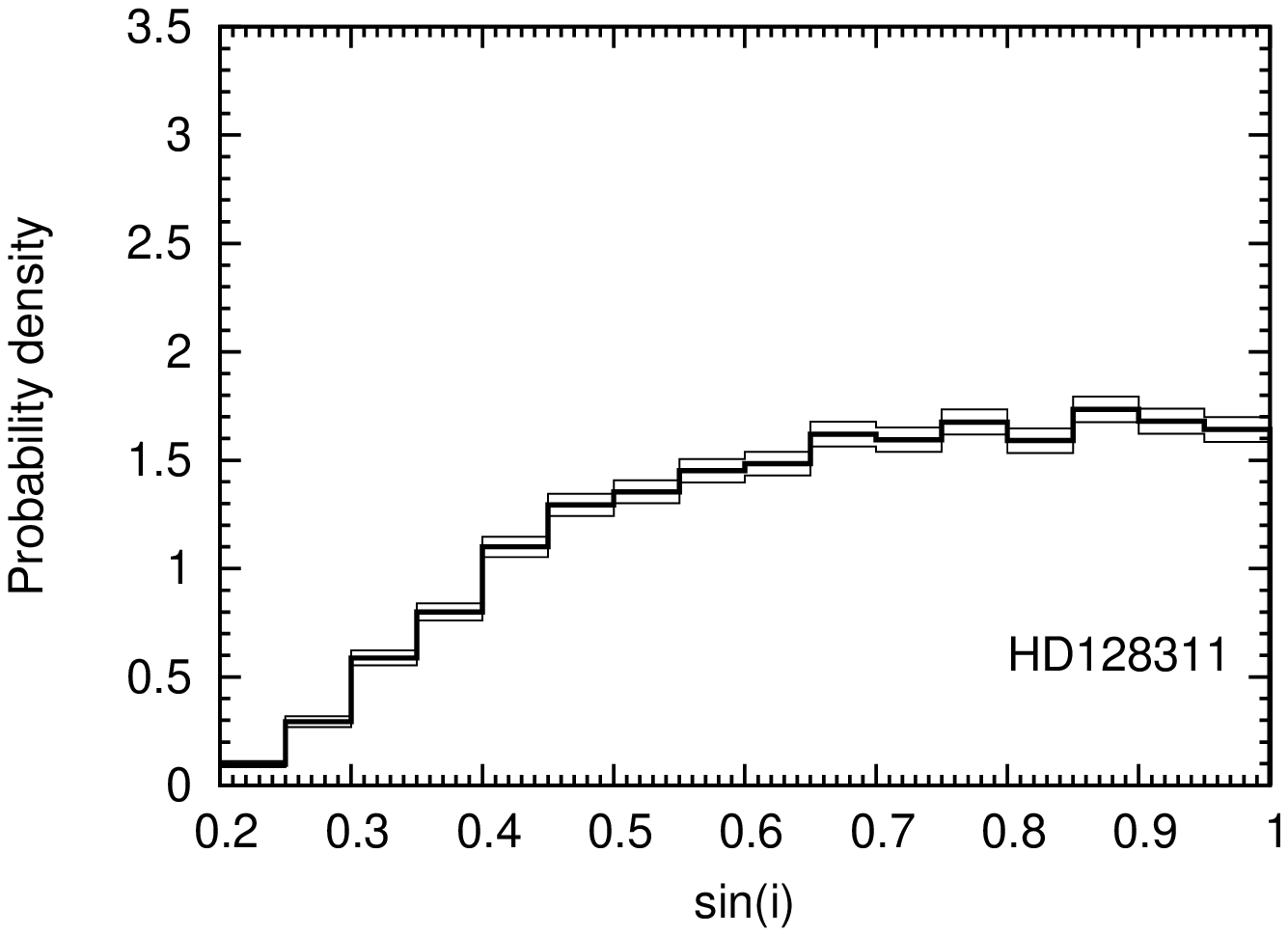}}%
\resizebox{55mm}{!}{\includegraphics{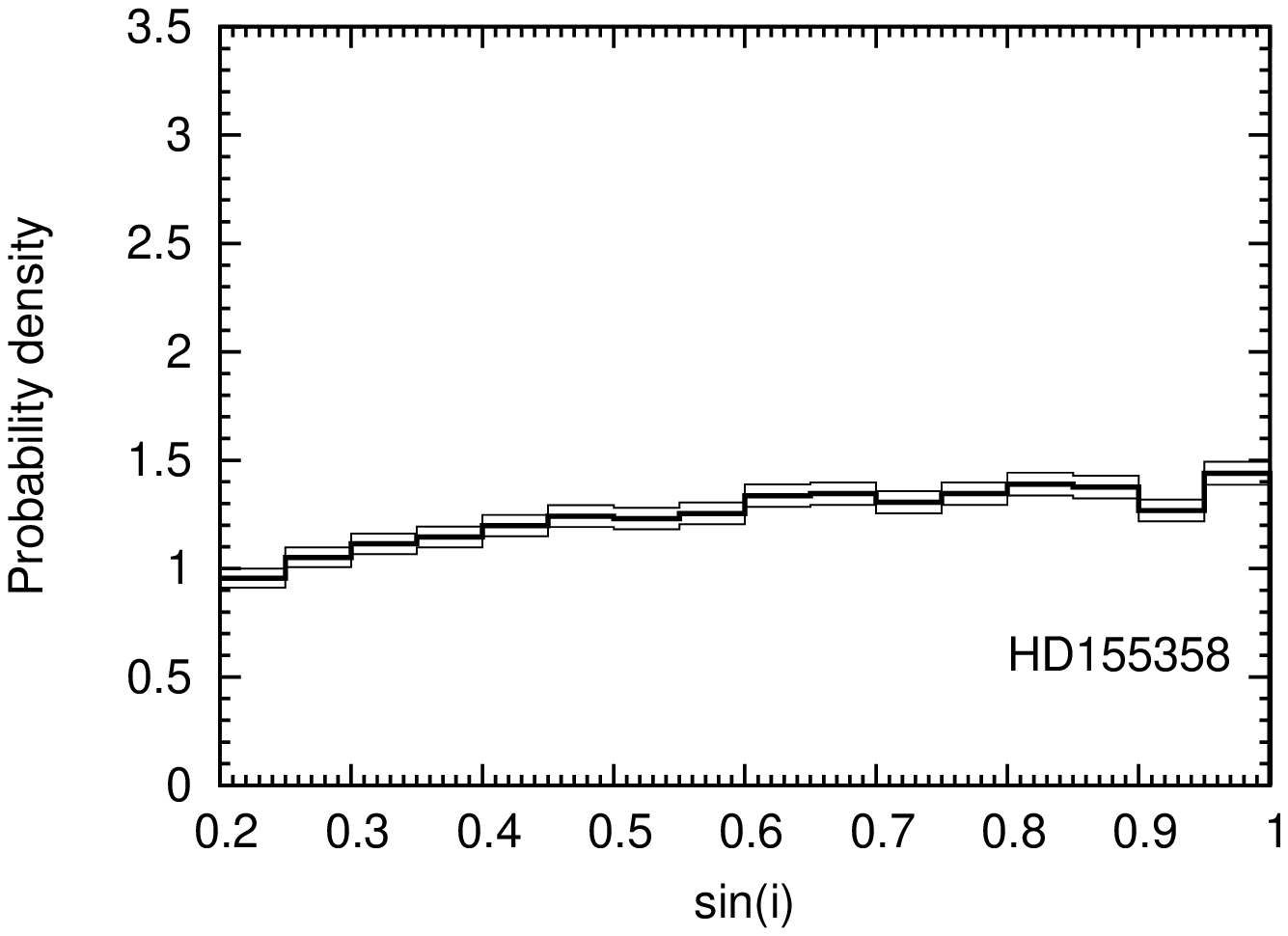}}
\end{center}
\caption{Probability distributions of the $\sin i$ orbital element
for the planetary systems HD~73526 (left panel), 
HD~128311 (middle panel) and HD~155358 (right panel) derived from
the radial velocity data available from the literature.}
\label{fig:sinidistributions}
\end{figure*}

\subsection{Generic functions}

Throughout this paper we were focusing on the analysis of radial velocity
variations of host stars in (multiple) planetary systems. However,
the results of the numerical $N$-body integrations can be exploited
in another aspects of binary and hierarchical stellar systems and extrasolar
planet studies -- including eclipsing binaries, transiting planets and 
planetary systems were astrometric information is also available. 
For instance, using the spatial coordinates $r_{km}$, 
one can estimate the moments of eclipses and transits and characterize
the shape of the light curves. Similarly, the velocities $w_{km}$
can directly be used to calculate the durations of the eclipses and
transits, since at the time of these events, the tangential 
acceleration of the transiting body is nearly zero, so the tangential
velocity computed from $w_{km}$ is a rather good approximation for
the reciprocal duration\footnote{In practice, total transit duration 
must be computed by taking into account the impact parameter
that is derived from the orbital inclination and the alignment of the
orbital ellipse. Therefore, the observed duration depends
on the $r_{km}$ coordinates as well.} Obviously,
using the capabilities of the program \texttt{lfit}, the knowledge of the
velocities and coordinates can be exploited in complex studies were 
simultaneous fits are performed on, for instance, RV and astrometric
data series \citep{mcarthur2010}.

First, let us write the independent
components of the barycentric coordinate and velocity in the form of 
\begin{eqnarray}
B_{{\rm r},m}^{(N,k)} & = & 
\frac{m_k r_{km}}{\mathcal{M}+\sum\limits_{i=1}^{N}m_i}\text{\hspace*{4ex}and}\label{eq:bynbodyk} \\
V_{{\rm r},m}^{(N,k)} & = & 
\frac{m_k w_{km}}{\mathcal{M}+\sum\limits_{i=1}^{N}m_i}.\label{eq:rvnbodyk} 
\end{eqnarray}
It can be shown that the projection of the barycentric coordinates 
to the line-of-sight is proportional to the light-time effects, thus
in the analysis of timing variations in eclipsing and/or transiting systems,
these corrections should also be taken into account. With the 
terms defined in \eqref{eq:rvnbodyk}, the observed radial velocity
of the host star can be written as
\begin{equation}
V_{\rm r}^{(N)}=\sum\limits_m\left(\sum\limits_{k=1}^{N} V_{{\rm r},m}^{(N,k)}\right)o_m.
\end{equation}
The independent components $V_{{\rm r},m}^{(N,k)}$ show the influence
of each body on the final observed radial velocity variations. However,
due to the mutual interactions, these functions are not strictly periodic.

In order to analyze the effects discussed above in multiple stellar
and planetary systems the module \texttt{nbrv.so} implements the functions 
\texttt{nbrv\_2g\_N()} and \texttt{nbrv\_3g\_N()} (where $\texttt{N}\equiv N$ 
is the number of planets). These functions have $4$ or $5$ additional
parameters\footnote{In the cases of $2$ and $3$ dimensional variants,
respectively.} comparing to the functions 
\texttt{nbrv\_2d\_N()} and \texttt{nbrv\_3d\_N()}. The 
first additional parameter is a mode flag, that is used as a selector 
between the $V_{\rm r}$, $B_{\rm r}$, $w$ and $r$ quantities. The
second additional parameter is the body index $k$ while the 
last $2$ or $3$ numbers represent the $(o_x,o_y)$ or $(o_x,o_y,o_z)$
vector. The value computed by the 
\texttt{nbrv\_2g\_N()} and \texttt{nbrv\_3g\_N()}
functions is a scalar product of the quantities $r_{km}$, $w_{km}$,
$B_{{\rm r},m}^{(N,k)}$ and $V_{{\rm r},m}^{(N,k)}$ and the $o_m$ 
vector. The comprehensive list of the modes implemented
in the functions \texttt{nbrv\_2g\_N()} and \texttt{nbrv\_3g\_N()}
can be found in Table~\ref{tab:genfunctmodes}.


\section{Applications}
\label{sec:applications}

In this section we describe two possible applications of the algorithms
presented in this paper. First, we show some examples related to 
the problem of orbital characterization. In the second part, we describe
in brief how optimal observation scheduling can be performed by employing
the \texttt{lfit} code and this implementation discussed above. 

\subsection{Orbital fits}
\label{sec:orbitalfits}

As a demonstration and application of the algorithm described
in Sec.~\ref{sec:algorithm}, we analyze the radial velocity
data for the multiple (double) planetary systems 
HD~73526 \citep{tinney2006},
HD~128311 \citep{vogt2005,wright2009a}
and 
HD~155358 \citep{cochran2007}
in the form as it is available from the literature. The objective of 
our test presented here is to constrain the orbital inclination assuming 
a planar model for these planetary systems. Coplanar orbits are expected
from modelling \citep{goldreich2004} and also confirmed by observations
\citep[Solar System or GJ~876, see][]{bean2009}. We note here that 
complex dynamical and stability investigations for multiple planetary systems 
can also be used to constrain or refine orbital elements 
\citep[see e.g.][]{ferrazmello2005}, imply alternative planetary configurations
\citep{gozdziewski2006} or rule out small inclinations 
\citep[in parallel with direct RV data modelling, see e.g.][]{laskar2009}.

We employed the method of Markov Chain Monte-Carlo 
\citep[MCMC, see e.g.][]{ford2005} in order to derive the 
best-fit orbital elements for these interacting planetary systems.
The initial conditions (orbital elements) were based on the data 
available in the literature and summarized in 
Table~\ref{tab:orbitalelements}. This table shows the 
orbital elements for a certain epoch ($E_0$) while the eccentricity,
argument of pericenter and the time of pericenter passage
have been converted to the spectroscopic orbital elements
discussed in Sec.~\ref{sec:speckepelements}. In all of the fits,
the model function
\begin{equation}
{\rm RV} = \gamma + (\sin i)V_{\rm r}^{(2)}\left(G\mathcal{M},\frac{\mathcal{K}_1}{\sin i},\mathbf{S}_1,\frac{\mathcal{K}_2}{\sin i},\mathbf{S}_2,t-E_0\right) 
\end{equation}
has been utilized. In the MCMC runs, the values of $\sin i$ have been
forced to be between $0.2$ and $1$ (note the probability that $\sin i$ of a 
randomly oriented orbit is less than $0.2$ is roughly 2\%). The
results of the fits are displayed in Table~\ref{tab:orbitalfits}.
The \emph{a posteriori} distributions of the $\sin i$ values
are shown in Fig~\ref{fig:sinidistributions}. As it is clear from the
plots, in the case of HD~73526, the orbital inclination is well defined
purely by the RV data, although the upper limit for $\sin i$ is not
constrained. Namely, $0.68 \lesssim \sin i$ for this system within 
$1$-$\sigma$, that is equivalent with $42^{\circ} \lesssim i$. 
For the other two
planetary systems, the available RV data do not provide 
a significant constraint for the inclination. 

An advantage of the knowledge of the partial derivatives of the 
model functions that the uncertainties of the fit parameters
can be estimated analytically involving the method of Fisher
matrix analysis \citep{finn1992,pal2009}. We have computed the 
uncertainties of the $\sin i$ parameters for these planetary
systems and obtained $\Delta(\sin i)=0.19$, $\Delta(\sin i)=3.87$
and $\Delta(\sin i)=6.14$ for 
HD~73526,  HD~128311 and HD~155358, respectively. The value
for HD~73526 well agrees with the result of the MCMC simulations,
while in the case of HD~128311 and HD~155358, these uncertainties
are definitely larger than $1$, indicating that the amount and/or
quality of available radial velocity data is not sufficient
for constraining the orbital inclination. Note that in general,
Fisher matrix analysis may underestimate the uncertainties if 
the probability distributions of the fitted variables cannot be 
approximated by Gaussian distributions. However, in our cases, where the
individual measurements are uncorrelated, their formal errors
are definitely smaller than the amplitude of the signal and the 
sampling of the model function is adequate (roughly homogeneous
for both periods), such a linear analysis yields reliable results. 

The results of this analysis have been compared with the results provided by
\texttt{Systemic Console} package \citep{meschiari2009} for the planetary 
system HD~73526. By taking into account the mutual perturbations,
the two applications yielded the same values for the best-fit 
parameters, however, the residual minimization procedure seemed to be
more sensitive for the initial parameters in the case of the \texttt{Systemic}
package. This is mainly due to the inadequate choices of the orbital 
elements\footnote{The \texttt{Systemic} package employs mean anomaly,
eccentricity and longitude of pericenter instead of mean longitude
and Lagrangian elements. A fit performed by \texttt{lfit}/\texttt{nbrv}
is also more unstable if the former set of orbital elements are used.}
The uncertainties derived by its built-in bootstrap method were 
roughly in the same magnitude, however, we were unable to derive 
such a large set of points that in the case of \texttt{lfit}/\texttt{nbrv}
since \texttt{Systemic} is slower by roughly two orders of magnitude. 

\subsection{Observation scheduling}

Since the program \texttt{lfit} is out-of-the-box capable to perform
analyses on arbitrary user input for which partial derivatives
are known and have an analytic property
(thus, it includes the usage of the \texttt{nbrv\_*()} functions as well),
these features can be exploited to optimize observation strategies
in order to derive more accurate orbital parameters. Recently,
\cite{ford2008} and \cite{baluev2008} gives methods with which
such strategies can be planned efficiently. As it is known
(see these papers), the computation of
all of the conditional probabilities, the expected information content
\citep{ford2008}, the D-optimal and L-optimal scheduling 
instances \citep{baluev2008,pal2009} requires the evaluation of the
covariance matrices in arbitrary instances. Since the program \texttt{lfit}
is capable for such an evaluation for arbitrary input functions,
with the aid of this program, these computations related to the 
optimal strategies can be performed as well without any serious difficulties.
By default, the program yields both the inverse of the information matrix
$\mathbf{Q}$ as defined also by \cite{baluev2008} and the goodness statistics
$\chi^2$, that and also appears in equations (13) -- (15) of \cite{ford2008}.
Additionally, \texttt{lfit} is capable to perform these kind of linear analyses
on arbitrary linear subspace of the parameter domain (i.e. parameters
in the orthogonal subspace are assumed to be fixed or known for independent
sources), therefore strategies can be built for optimizing various 
combinations of orbital parameters. 


\section{Summary}

In this paper we described an algorithm based on the 
Lie-integration method that efficiently computes the parametric derivatives
of radial velocity model functions for multiple planetary systems
when the planet-planet interactions are also taken into account. 
The analysis of these systems yields more accurate constrains
for planetary masses since the orbital and mutual inclinations 
can also be derived if precise radial velocity data are available.
Additionally, the presented analytic formulae and integration
method aid to plan observation schedules in order to optimize the 
telescope time utilization in order to detect planetary perturbations.


\section*{Acknowledgments}

The author would thank L\'aszl\'o Szabados for the careful 
reading of the draft and for suggestions of improvements and 
the anonymous referee for further ideas and the careful proofreading.
The author would also thank for the numerous discussions with colleagues
attended in the Fifth Austrian-Hungarian Workshop on Trojans and Related Topics
(Vienna, 2010) that has also been improved the quality of the code.
This work has been supported by the scholarship of the Doctoral School 
of the E\"otv\"os University and also in part by ESA grant PECS~98073. 


{}


\onecolumn
\appendix

\section{Motion in a reference frame fixed to one of the bodies}
\label{appendix:fixednb}

Recalling \cite{pal2007}, Appendix C, the recurrence relations
for the $N$-body problem around a fixed center can be written as
\begin{eqnarray}
L^{n+1}r_{im} & = & L^nw_{im}, \label{fixnbrec}\\
L^{n}A_{ijm} & = & L^nr_{im}-L^nr_{jm},  \\
L^{n}B_{ijm} & = & L^nw_{im}-L^nw_{jm},  \\
L^{n}\Lambda_{i} & = & \sum\limits_{k=0}^{n}\binom{n}{k}L^kr_{im}L^{n-k}w_{im},  \\
L^{n}\Lambda_{ij} & = & \sum\limits_{k=0}^{n}\binom{n}{k}L^kA_{ijm}L^{n-k}B_{ijm},  \\
L^{n+1}w_{im} & = & -G(\mathcal{M}+m_i)\sum\limits_{k=0}^n\binom{n}{k}L^k\phi_iL^{n-k}r_{im} - G\sum\limits_{j=1, j\ne i}^{N}m_j\sum\limits_{k=0}^n\binom{n}{k}\left[L^k\phi_{ij}L^{n-k}A_{ijm}+L^k\phi_jL^{n-k}r_{jm}\right], \\
L^{n+1}\phi_{i} & = & \rho_{i}^{-2}\sum\limits_{k=0}^{n}F_{nk}L^{n-k}\phi_{i}L^k\Lambda_{i},  \\
L^{n+1}\phi_{ij} & = & \rho_{ij}^{-2}\sum\limits_{k=0}^{n}F_{nk}L^{n-k}\phi_{ij}L^k\Lambda_{ij}, \label{fixnbrec2}
\end{eqnarray}
where $F_{nk}=(-3)\binom{n}{k}+(-2)\binom{n}{k+1}$.
Since the masses (both $\mathcal{M}$ and $m_i$) are constants,
the Lie-derivatives of these are simply 
$L^{n+1}\mathcal{M}=L^{n+1}m_i=0$ for $0\le n$.
Let us denote the linearized of 
$r_{im}$, $w_{im}$, $m_i$ and $\mathcal{M}$ by 
$\xi_{im}$, $\eta_{im}$, $\mathfrak{m}_i$ and $\mathfrak{M}$, respectively. 
The vectors $\Xi$ and $\mathcal{D}$ must be extended with
the linearized variables $\mathfrak{m}_i$ and $\mathfrak{M}$,
namely
\begin{equation}
\hat\Xi  = \big( \{\xi_{kp}\}, \{\eta_{kp}\}, \{\mathfrak{m}_k\}, \mathfrak{M} \big)
\hspace*{3ex}\text{and}\hspace*{3ex}
\hat{\mathcal{D}} = \left( 	\left\{\frac{\partial}{\partial r_{kp}}\right\}, 
				\left\{\frac{\partial}{\partial w_{kp}}\right\}, 
				\left\{\frac{\partial}{\partial m_k}\right\}, 
				\frac{\partial}{\partial\mathcal{M}} \right)
\end{equation}
and hence
\begin{equation}
\hat\Xi\cdot\hat{\mathcal{D}} = 
	\left(\sum\limits_{k,p}\xi_{kp}\frac{\partial}{\partial r_{kp}}\right)+
	\left(\sum\limits_{k,p}\eta_{kp}\frac{\partial}{\partial w_{kp}}\right)+
	\left(\sum\limits_{k}\mathfrak{m}_k\frac{\partial}{\partial m_k}\right)+
	\mathfrak{M}\frac{\partial}{\partial\mathcal{M}}.
\end{equation}
It can easily be shown that the complete set of linearized equations
extended with the variables $\mathfrak{m}_i$ and $\mathfrak{M}$ are 
\begin{eqnarray}
L^{n+1}\xi_{im} & = & L^n\eta_{im}, \label{fixnblinrec}\\
L^{n}\alpha_{ijm} & = & L^n\xi_{im}-L^n\xi_{jm},  \\
L^{n}\beta_{ijm} & = & L^n\eta_{im}-L^n\eta_{jm}, \\
\hat\Xi\cdot\hat{\mathcal{D}}L^{n}\Lambda_{i} & = & \sum\limits_{k=0}^{n}\binom{n}{k}\left(L^k\xi_{im}L^{n-k}w_{im}+L^kr_{im}L^{n-k}\eta_{im}\right), \\
\hat\Xi\cdot\hat{\mathcal{D}}L^{n}\Lambda_{ij} & = & \sum\limits_{k=0}^{n}\binom{n}{k}\left(L^k\alpha_{ijm}L^{n-k}B_{ijm}+L^kA_{ijm}L^{n-k}\beta_{ijm}\right),  \\
L^{n+1}\eta_{im} & = & -G(\mathfrak{M}+\mathfrak{m}_i)\sum\limits_{k=0}^n\binom{n}{k}L^k\phi_iL^{n-k}r_{im} - G\sum\limits_{j=1, j\ne i}^{N}\mathfrak{m}_j\sum\limits_{k=0}^n\binom{n}{k}\left[L^k\phi_{ij}L^{n-k}A_{ijm}+L^k\phi_jL^{n-k}r_{jm}\right] - \\
		& & -G(\mathcal{M}+m_i)\sum\limits_{k=0}^n\binom{n}{k}\left[(\hat\Xi\cdot\hat{\mathcal{D}}L^k\phi_i)L^{n-k}r_{im}+L^k\phi_iL^{n-k}\xi_{im}\right] - \nonumber \\
                & & - G\sum\limits_{j=1, j\ne i}^{N}m_j\sum\limits_{k=0}^n\binom{n}{k}\left[(\hat\Xi\cdot\hat{\mathcal{D}}L^k\phi_{ij})L^{n-k}A_{ijm}+L^k\phi_{ij}L^{n-k}\alpha_{ijm}+(\hat\Xi\cdot\hat{\mathcal{D}}L^k\phi_j)L^{n-k}r_{jm}+L^k\phi_jL^{n-k}\xi_{jm}\right], \nonumber \\
\hat\Xi\cdot\hat{\mathcal{D}}L^{n+1}\phi_{i} & = & -2\rho_{i}^{-2}\xi_{im}r_{im}L^{n+1}\phi_{i}+ \rho_{i}^{-2}\sum\limits_{k=0}^{n}F_{nk}\left[(\hat\Xi\cdot\hat{\mathcal{D}}L^{n-k}\phi_{i})L^k\Lambda_{i}+L^{n-k}\phi_{i}(\hat\Xi\cdot\hat{\mathcal{D}}L^k\Lambda_{i})\right],  \\
\hat\Xi\cdot\hat{\mathcal{D}}L^{n+1}\phi_{ij} & = & -2\rho_{ij}^{-2}\alpha_{ijm}A_{ijm}L^{n+1}\phi_{ij} + \rho_{ij}^{-2}\sum\limits_{k=0}^{n}F_{nk}\left[(\hat\Xi\cdot\hat{\mathcal{D}}L^{n-k}\phi_{ij})L^k\Lambda_{ij}+L^{n-k}\phi_{ij}(\hat\Xi\cdot\hat{\mathcal{D}}L^k\Lambda_{ij})\right]. \label{fixnblinrec2}
\end{eqnarray}
Obviously, the auxiliary variables 
$S^{[n]}_{im}$, $S^{[n]}_{ijm}$, $\Sigma^{[n]}_{im}$ and $\Sigma^{[n]}_{ijm}$
can be introduced as well \citep[see][Appendix D]{pal2007}, in order to optimize
the evaluation of equations (\ref{fixnblinrec})~--~(\ref{fixnblinrec2}).

\section{Partial derivatives of the coordinates and velocities}
\label{appendix:partial}

The computation of \eqref{eq:qpartial2} requires the partial derivatives
of the initial coordinates and velocities with respect to the
initial orbital elements. Let us write the initial normalized 
coordinates and velocities as
\begin{eqnarray}
\binom{\xi}{\eta} & = & 
	\binom{c}{s}+\frac{p}{1+J}\binom{+h}{-k}-\binom{k}{h}, \\
\binom{\xi'}{\eta'}\equiv \frac{\partial}{\partial\lambda}\binom{\xi}{\eta} & = & 
	\frac{1}{1-q}\left[\binom{-s}{+c}+\frac{q}{1+J}\binom{+h}{-k}\right]
\end{eqnarray}
The normalized coordinates and velocities do not depend on the 
semimajor axis and the mean motion, therefore these quantities
are only functions of the mean longitude $\lambda$ and the 
Lagrangian orbital elements $(k,h)$. In the above equations,
$p\equiv\eop(\lambda,k,h)$, $q\equiv\eoq(\lambda,k,h)$,
$c\equiv\cos(p+\lambda)$, $s\equiv\sin(p+\lambda)$ and
the quantity $J$ is defined as $J=\sqrt{1-e^2}=\sqrt{1-k^2-h^2}$.

Thus, the partial derivatives of the mass parameter
$Gm$, coordinates $(x,y)$ and velocities $(\dot x,\dot y)$ with respect to the
central mass parameter $G\mathcal{M}$ and the 
spectroscopic orbital elements -- normalized semi-amplitude $\mathcal{K}$,
mean motion $n$, mean longitude $\lambda$ and the Lagrangian
orbital elements $(k,h)$ -- are then
\begin{equation}
\frac{\partial(Gm,x,y,\dot x,\dot y)}{\partial(G\mathcal{M},\mathcal{K},n,\lambda,k,h)}=
\begin{pmatrix}
\dfrac{2m}{3\mathcal{M}+m} &
\dfrac{3\mathcal{K}^2(\mathcal{M}+m)^3}{n  (3\mathcal{M}+m)m^2} & 
-\dfrac{\mathcal{K}^3(\mathcal{M}+m)^3}{n^2(3\mathcal{M}+m)m^2} &
0 &
0 &
0 \\[3mm]

\dfrac{\partial a}{\partial(G\mathcal{M})}\xi & 
\dfrac{\partial a}{\partial \mathcal{K}}\xi & 
\dfrac{\partial a}{\partial n}\xi &
a\xi' &
a\dfrac{\partial \xi}{\partial k} &
a\dfrac{\partial \xi}{\partial h} \\[3mm]

\dfrac{\partial a}{\partial(G\mathcal{M})}\eta & 
\dfrac{\partial a}{\partial \mathcal{K}}\eta & 
\dfrac{\partial a}{\partial n}\eta &
a\eta' &
a\dfrac{\partial \eta}{\partial k} &
a\dfrac{\partial \eta}{\partial h} \\[3mm]

\dfrac{\partial a}{\partial(G\mathcal{M})}n\xi'& 
\dfrac{\partial a}{\partial \mathcal{K}}n\xi' & 
\left(a+n\dfrac{\partial a}{\partial n}\right)\xi' &
an\dfrac{\partial\xi'}{\partial\lambda} &
an\dfrac{\partial\xi'}{\partial k} &
an\dfrac{\partial\xi'}{\partial h} \\[3mm]

\dfrac{\partial a}{\partial(G\mathcal{M})}n\eta'& 
\dfrac{\partial a}{\partial \mathcal{K}}n\eta' & 
\left(a+n\dfrac{\partial a}{\partial n}\right)\eta' &
an\dfrac{\partial\eta'}{\partial\lambda} &
an\dfrac{\partial\eta'}{\partial k} &
an\dfrac{\partial\eta'}{\partial h} 

\end{pmatrix}.
\end{equation}
Here $(\xi',\eta')\equiv\dfrac{\partial(\xi,\eta)}{\partial\lambda}$ and
\begin{eqnarray}
\frac{\partial a}{\partial(G\mathcal{M})}	& = & \frac{a}{G(3\mathcal{M}+m)} \\
\frac{\partial a}{\partial \mathcal{K}}	& = & \frac{3a\mathcal{K}^2(\mathcal{M}+m)^2}{3nGm^2(3\mathcal{M}+m)} \\
\frac{\partial a}{\partial n}	& = & -\frac{2a}{3n}-\dfrac{a\mathcal{K}^3(\mathcal{M}+m)^2}{3n^2Gm^2(3\mathcal{M}+m)}.
\end{eqnarray}
The partial derivatives of the normalized coordinates $(\xi,\eta)$ and
the normalized velocities $(\xi',\eta')$ with respect to the
orbital elements $(\lambda,k,h)$ are the following:
\begin{eqnarray}
\frac{\partial(\xi,\eta)}{\partial(k,h)} & = & 
	\frac{1}{1-q}\left[
	-\begin{pmatrix}
	s^2 & -sc \\ -sc & c^2 
	\end{pmatrix}
	+\frac{1}{1+J}\begin{pmatrix}
	sh & -ch \\ -sk & ck
	\end{pmatrix}
	\right]+
	\frac{p}{J(1+J)^2}\begin{pmatrix}
	kh & h^2 \\ -k^2 & -kh 
	\end{pmatrix}
	+\begin{pmatrix}
	-1 & \dfrac{p}{1+J} \\ \dfrac{-p}{1+J} & -1
	\end{pmatrix}, \\
\frac{\partial}{\partial\lambda}\binom{\xi'}{\eta'} & = & 
	\frac{1}{(1-q)^3}\left[-\binom{c}{s}-\frac{p}{1+J}\binom{+h}{-k}+\binom{k}{h}\right], \\
\frac{\partial(\xi',\eta')}{\partial(k,h)} & = & 
	\frac{1}{(1-q)^3}
	\begin{pmatrix}
		-2sc   +ks-qsc +\dfrac{h(c-k)}{1+J} & c^2-s^2+hs-qc^2+\dfrac{h(s-h)}{1+J} \\[2mm]
		c^2-s^2-ck+qs^2-\dfrac{k(c-k)}{1+J} & 2sc    -ch-qsc -\dfrac{k(s-h)}{1+J} 
	\end{pmatrix} + \nonumber \\
& & 	+\frac{q}{(1-q)J(1+J)^{2}}
	\begin{pmatrix}
		kh & h^2 \\
		-k^2 & -kh
	\end{pmatrix} + 
	\frac{q}{(1-q)(1+J)}
	\begin{pmatrix}
		0 & 1 \\
		-1 & 0 
	\end{pmatrix}.
\end{eqnarray}
\bsp

\label{lastpage}


\begin{thebibliography}{99}
\bibitem[\protect\citeauthoryear{Bakos et al.}{2009}]{bakos2009}
Bakos, G. \'A. et al.
2009, ApJ, 707, 446

\bibitem[\protect\citeauthoryear{Bean \& Seifahrt}{2009}]{bean2009}
Bean, J. L. \& Seifahrt, A.
2009, A\&A, 496, 249

\bibitem[\protect\citeauthoryear{Benedict et al.}{2010}]{benedict2010}
Benedict, G. F., McArthur, B. E., Bean, J. L.; Barnes, R., Harrison, T. E., Hatzes, A., Martioli, E. \& Nelan, E. P.
2010, AJ, 139, 1844

\bibitem[\protect\citeauthoryear{Baluev}{2008}]{baluev2008}
Baluev, R. V.
2008, MNRAS, 389, 1375

\bibitem[\protect\citeauthoryear{Borkovits et al.}{2003}]{borkovits2003}
Borkovits, T., \'Erdi, B., Forg\'acs-Dajka, E. \& Kov\'acs, T.
2003, A\&A, 398, 1091

\bibitem[\protect\citeauthoryear{Cochran et al.}{2007}]{cochran2007}
Cochran, W. D., Endl, M., Wittenmyer, R. A. \& Bean, J. L.
2007, ApJ, 665, 1407

\bibitem[\protect\citeauthoryear{Eggl \& Dvorak}{2010}]{eggl2010}
Eggl, S. \& Dvorak, R.
2010, Lecture Notes in Physics: ``An Introduction to Common Numerical Integration Codes Used in Dynamical Astronomy'', eds. J. Souchay and R. Dvorak, Vol. 790, Springer

\bibitem[\protect\citeauthoryear{Ferraz-Mello, Michtchenko, \& Beaug\'e}{2005}]{ferrazmello2005}
Ferraz-Mello, S., Michtchenko, T. A. \& Beaug\'e, C.
2005, ApJ, 621, 473

\bibitem[\protect\citeauthoryear{Finn}{1992}]{finn1992}
Finn, L. S.
1992, Phys.~Rev.~D, 46, 5236

\bibitem[\protect\citeauthoryear{Ford}{2005}]{ford2005}
Ford, E. B. 
2005, AJ, 129, 1706

\bibitem[\protect\citeauthoryear{Ford}{2008}]{ford2008}
Ford, E. B. 
2008, AJ, 135, 1008

\bibitem[\protect\citeauthoryear{Goldreich, Lithwick \& Sari}{2004}]{goldreich2004}
Goldreich, P., Lithwick, Y. \& Sari, R.
2004, ApJ, 614, 497

\bibitem[\protect\citeauthoryear{Go\'zdziewski \& Konacki}{2006}]{gozdziewski2006}
Go\'zdziewski, K. \& Konacki, M.
2006, ApJ, 647, 573


\bibitem[\protect\citeauthoryear{Gr\"obner \& Knapp}{1967}]{grobner1967} 
Gr\"obner, W. \& Knapp, H.
1967, "Contributions to the Method of Lie-Series",
Bibliographisches Institut, Mannheim

\bibitem[\protect\citeauthoryear{Hanslmeier \& Dvorak}{1984}]{hanslmeier1984} 
Hanslmeier, A. \& Dvorak, R.
1984, A\&A, 132, 203

\bibitem[\protect\citeauthoryear{Kane et al.}{2009}]{kane2009}
Kane, S. R., Mahadevan, S., von Braun, K., Laughlin, G. \& Ciardi, D. R.
2009, PASP, 121, 1386

\bibitem[\protect\citeauthoryear{Laskar \& Correia}{2009}]{laskar2009} 
Laskar, J. \& Correia, A. C. M.
2009, A\&A, 496, 5

\bibitem[\protect\citeauthoryear{Laughlin \& Chambers}{2001}]{laughlin2001}
Laughlin, G. \& Chambers, J. E.
2001, ApJ, 551, 109

\bibitem[\protect\citeauthoryear{McArthur et al.}{2010}]{mcarthur2010}
McArthur, B. E., Benedict, G. F., Barnes, R., Martioli, E., Korzennik, S.; Nelan, E. \& Butler, R. P.
2010, ApJ, 715, 1203

\bibitem[\protect\citeauthoryear{Marois et al.}{2008}]{marois2008}
Marois, C., Macintosh, B., Barman, T., Zuckerman, B., Song, I., Patience, J., Lafreni\`ere, D. \& Doyon, R.
2008, Science, 322, 1348

\bibitem[\protect\citeauthoryear{Meschiari et al.}{2009}]{meschiari2009}
Meschiari, S., Wolf, A., Rivera, E., Laughlin, G., Vogt, S. \& Butler, P. 
2009, PASP, 121, 1016

\bibitem[\protect\citeauthoryear{Murray \& Dermott}{1999}]{murray1999}
Murray, C. D. \& Dermott, S. F.
1999, Solar System Dynamics, Cambridge Univ. Press, Cambridge

\bibitem[\protect\citeauthoryear{P\'al \& S\"uli}{2007}]{pal2007}
P\'al, A. \& S\"uli, \'A.
2007, MNRAS, 381, 1515

\bibitem[\protect\citeauthoryear{P\'al}{2009a}]{pal2009}
P\'al, A.
2009a, MNRAS, 396, 1737

\bibitem[\protect\citeauthoryear{P\'al}{2009b}]{pal2009b}
P\'al, A.
2009b, PhD thesis (arXiv:0906.3486) 

\bibitem[\protect\citeauthoryear{P\'al et al.}{2010}]{pal2010}
P\'al, A. et al. 
2010, MNRAS, 401, 2665

\bibitem[\protect\citeauthoryear{Press et al.}{1992}]{press1992}
Press, W. H., Teukolsky, S. A., Vetterling, W.T. \& Flannery, B.P.
1992, Numerical  Recipes in C: the art of scientific computing, Second Edition, Cambridge University Press

\bibitem[\protect\citeauthoryear{Tinney et al.}{2006}]{tinney2006}
Tinney, C. G., Butler, R. P., Marcy, G. W., Jones, H. R. A., Laughlin, G., Carter, B. D., Bailey, J. A. \& O'Toole, S.
2006, ApJ, 647. 594

\bibitem[\protect\citeauthoryear{Vogt et al.}{2005}]{vogt2005}
Vogt, S. S., Butler, R. P., Marcy, G. W., Fischer, D. A., Henry, G. W., Laughlin, G., Wright, J. T. \& Johnson, J. A.
2005, ApJ, 632, 638

\bibitem[\protect\citeauthoryear{Wright et al.}{2009}]{wright2009a}
Wright, J. T., Upadhyay, S., Marcy, G. W., Fischer, D. A., Ford, E. B. \& Johnson, J. A.
2009, ApJ, 693, 1084

\bibitem[\protect\citeauthoryear{Wright \& Howard}{2009}]{wright2009b}
Wright, J. T. \& Howard, A. W.
2009, ApJS, 182, 205
\end{thebibliography}
\end{document}